\documentclass[journal,draftcls,onecolumn,12pt,twoside]{IEEEtran}
\normalsize

\usepackage{lineno}
\usepackage{cite}
\ifCLASSINFOpdf
\else
\fi

\usepackage{amsmath,amsfonts}
\usepackage{algorithmic}
\usepackage{algorithm}
\usepackage{array}
\usepackage[caption=false,font=normalsize,labelfont=sf,textfont=sf]{subfig}
\usepackage{textcomp}
\usepackage{stfloats}
\usepackage{url}
\usepackage{verbatim}
\usepackage{graphicx}
\usepackage{cite}
\usepackage{xcolor}
\usepackage{multirow}

\newtheorem{lemma}{Lemma}

\newenvironment{proof}{\par{\it{Proof:}}\ }{\hfill $\square$\par}

\usepackage{algorithm}
\usepackage{algorithmic}
\usepackage{float}
\usepackage{lipsum}
\usepackage{xcolor}
\newcommand{\F}{\mathbb{F}}


\newtheorem{aq}{Author Query/Comment}

\newcommand{\baq}{\begin{aq}}
\newcommand{\eaq}{\end{aq}}

\begin{document}
\title{Computing Vanishing Polynomials with Fast Walsh-Hadamard Transforms}

\author{
	Nianqi Tang, Yunghsiang S. Han~\IEEEmembership{Fellow,~IEEE}, Chao Chen
	\thanks{
		Nianqi Tang (724973040@qq.com) is with the Theo End (Nanjing) Computing Technology Co., Ltd., China.  Yunghsiang S. Han (yunghsiangh@gmail.com) is with the School of Computing and Information Technology, Great Bay University, Dongguan, China. Chao Chen (cchen@xidian.edu.cn) is with the State Key Lab of ISN, Xidian University, Xi'an, China.
		}% <-this % stops a space
%\thanks{Manuscript received October 18, 2019; revised December 05, 2019.}
%%}
}

\maketitle
\begin{abstract}
This paper proposes an algorithm for computing vanishing polynomials with fast Walsh-Hadamard transforms over the field $GF(2^m)$, whose computational complexity is $O(2^m\log(2^{\tau}))$, where $\tau$ is smallest non-negative integer such that $2^{\tau}$ is larger than or equal to the degree of desired vanishing polynomial. The new algorithm outperforms the Didier's method whose computational complexity is $O(2^m\log(2^{m}))$.
\end{abstract}

% \fi
\IEEEpeerreviewmaketitle

\section{Introduction}
Let $\F = GF(2^m)$ be a finite field. Given a subset $E\subseteq \F$, the vanishing polynomial $\Lambda(x)$ with respect to $E$ is defined as
\begin{equation*}
	\Lambda(x) = \prod_{\alpha\in E}(x - \alpha).
\end{equation*}
If $E$ is an empty set, $\Lambda(x)$ is defined to be $1$.
The problem of computing vanish polynomials is exactly the problem of computing $\Lambda(x)$ for a given set $E$.
Without loss of generality, we assume that $|E| > 0$ hereafter. 

Computing vanishing polynomials is important in many applications. For example, in the erasure decoding of Reed-Solomon (RS) codes, the decoder first computes the vanishing polynomial with respect to the set of erasure locations, then figures out the corresponding erasure values. More details are referred to \cite{Lin20161}. 

To the best of the authors' knowledge, the algorithm with the lowest computational complexity for computing vanishing polynomial to date is the Didier's method, see \cite{Didier2009}. It makes use of $2^m$-point fast Walsh-Hadamard transform (FWHT) and has a complexity of $O(2^m\log(2^{m}))$. In this paper, we show that the problem of computing vanishing polynomials can be solved by using $2^{\tau}$-point FWHT, where $\tau$ is smallest non-negative integer such that $2^{\tau}\geq |E|$. This yields a new algorithm whose complexity is $O(2^m\log(2^{\tau}))$.

\section{New Algorithm}
This section derives the new algorithm.

Let $\{v_0, v_1, \dots, v_{m - 1}\}$ be a basis of the linear space $\F = GF(2^m)$ over $GF(2)$ and let
\begin{equation*}
	\omega_i = i_0\cdot v_0 + i_1\cdot v_1 + i_{m - 1} \cdot v_{m - 1}, i = 0, 1, \dots, 2^m - 1,
\end{equation*}
where $i_j$ is the binary representation of $i$. Clearly, the field $\F$ can be written as
\begin{equation*}
	\F = \{\omega_0, \omega_1, \dots, \omega_{2^m - 1}\}.
\end{equation*}
For $0\leq\tau \leq m$, the subset
\begin{equation*}
	K_{\tau} = \{\omega_0, \omega_1, \dots, \omega_{2^{\tau} - 1}\}
\end{equation*}
is an additive subgroup of $\F$ and the subset
\begin{equation*}
	P_j = \{\omega_{j\cdot 2^{\tau}}, \omega_{j\cdot 2^{\tau} + 1}, \dots, \omega_{j\cdot 2^{\tau} + 2^{\tau} - 1}\}
\end{equation*}
is a coset of $K_{\tau}$. The field $\F$ can also be written as
\begin{equation*}
	\F  = \{P_0, P_1, \dots, P_{\eta - 1}\},
\end{equation*}
where $\eta = 2^{m - \tau}$ and $P_0 = K_{\tau}$.

Define
\begin{equation*}
	E_j = E\cap P_j, \Lambda_{j}(x) = \prod_{\omega_{j\cdot 2^{\tau} + l}\in E_j}(x - \omega_{j\cdot 2^{\tau} + l}). 
\end{equation*}
Given $E_j$, computing $\Lambda_{j}(x)$ can be seen as a special case of computing vanishing polynomials problem, which satisfies that $E_j$ is contained in the coset $P_j$. For this specific problem, we have the following lemma:
\begin{lemma}
	Given $E_j$, the polynomial $\Lambda_j(x)$ can be obtained by $2^{\tau}$-point FWHT, and the computational complexity is $O(2^{\tau}\log(2^{\tau}))$.
	
	\begin{proof}
		The elements of $P_j$ can be written as
		\begin{equation*}
			\omega_{j\cdot 2^{\tau}} + \omega_0, \omega_{j\cdot 2^{\tau}} + \omega_1, \dots, \omega_{j\cdot 2^{\tau}} + \omega_{2^{\tau} - 1}, 
		\end{equation*}
		where $\omega_{j\cdot 2^{\tau}}\in \F$.
		
		For $\omega_{j\cdot 2^{\tau} + r}\in P_j\setminus E_j$, we have
		\begin{equation*}
			\Lambda_j(\omega_{j\cdot 2^{\tau} + r}) = \prod_{\omega_{j\cdot 2^{\tau} + l}\in E_j}(\omega_{j\cdot 2^{\tau} + r} - \omega_{j\cdot 2^{\tau} + l}).
		\end{equation*}
		
		Define $R_{\omega_{j\cdot 2^{\tau} + l}}$ as
		\begin{equation*}
			R_{\omega_{j\cdot 2^{\tau} + l}} =
			\begin{cases}
				1,\ \omega_{j\cdot 2^{\tau} + l}\in E_j,\\
				0,\ \omega_{j\cdot 2^{\tau} + l}\in P_j\setminus E_j.
			\end{cases}
		\end{equation*}
		One has
		\begin{equation*}
			\Lambda_j(x) = \prod_{\omega_{j\cdot 2^{\tau} + l}\in P_j}(x - \omega_{j\cdot 2^{\tau} + l})^{R_{\omega_{j\cdot 2^{\tau} + l}}}.
		\end{equation*}
		Then for $\omega_{j\cdot 2^{\tau} + r}\in P_j\setminus E_j$,
		\begin{equation*}
			\Lambda_j(\omega_{j\cdot 2^{\tau} + r}) = \prod_{\omega_{j\cdot 2^{\tau} + l}\in P_j}(\omega_{j\cdot 2^{\tau} + r} - \omega_{j\cdot 2^{\tau} + l})^{R_{\omega_{j\cdot 2^{\tau} + l}}}.
		\end{equation*}
		Taking the logarithm of both sides of the above equation, we have
		\begin{equation}\label{eq:logical_convolution}
			\log(\Lambda_j(\omega_{j\cdot 2^{\tau} + r})) = \sum_{\omega_{j\cdot 2^{\tau} + l}\in P_j}R_{\omega_{j\cdot 2^{\tau} + l}}\log(\omega_{j\cdot 2^{\tau} + r} - \omega_{j\cdot 2^{\tau} + l}).
		\end{equation}
		For elements $\omega_{j\cdot 2^{\tau} + r}, \omega_{j\cdot 2^{\tau} + l}\in P_j$, one can verify that
		\begin{align*}
			\omega_{j\cdot 2^{\tau} + r} -  \omega_{j\cdot 2^{\tau} + l} &= \omega_{j\cdot 2^{\tau}} + \omega_r - (\omega_{j\cdot 2^{\tau}} + \omega_l)\\
			&= \omega_r - \omega_l\\
			&= \omega_r + \omega_l\\
			&= \omega_{r\oplus l}.
		\end{align*}
		If we write
		\begin{equation*}
			T_l = R_{\omega_{j\cdot 2^{\tau} + l}}, L_l = \log(\omega_l),
		\end{equation*}
		The equation \eqref{eq:logical_convolution} becomes
		\begin{equation}\label{eq:logical_convolution_v1}
			\log(\Lambda_j(\omega_{j\cdot 2^{\tau} + r})) = \sum_{l = 0}^{2^{\tau} - 1}T_{l}\cdot L_{r\oplus l}.
		\end{equation}
		Hence, \eqref{eq:logical_convolution_v1} is logical convolution, which can be computed by $2^{\tau}$-point FWHT. Clearly, the complexity of computing \eqref{eq:logical_convolution_v1} is $O(2^{\tau}\log(2^{\tau}))$.
	\end{proof}
\end{lemma}

We write
\begin{equation*}
	\mathbf{T} = (T_0, T_1, \dots, T_{2^{\tau} - 1}), \mathbf{L} = (L_0, L_1, \dots, L_{2^{\tau} - 1}).
\end{equation*}
and let
\begin{equation*}
	\mathbf{\hat{T}} = \text{FWHT}_{2^{\tau}}(\mathbf{T}), \mathbf{\hat{L}} = \text{FWHT}_{2^{\tau}}(\mathbf{L})
\end{equation*}
represent the $2^{\tau}$-point FWHT transforms of $\mathbf{T}$ and $\mathbf{L}$, respectively.
Algorithm \ref{alg:coset} describes the proposed method of computing $\Lambda_{j}(x)$. It should be mentioned that, the vector $\text{FWHT}_{2^{\tau}}(\mathbf{L})$ can be precomputed for each possible $\tau$, which requires $O(2^m)$ storage space. In addition, when $|E_j| = 2^{\tau}$, $\Lambda_{j}(x) = s_{\tau}(x) - s_{\tau}(\omega_{j\cdot2^{\tau}})$, where $s_{\tau}$ is the subspace polynomial, see \cite{Lin20161} for details.

\begin{algorithm}[h]
	\caption{Computing $\Lambda_j(x)$}
	\label{alg:coset}
	\begin{algorithmic}[1]
		\REQUIRE $E_j$, $\tau$.
		\ENSURE $\Lambda_j(x)$.
		\IF{$|E_j| = 0$}
		\STATE $\Lambda_j(x) = 1$.
		\ELSIF{$|E_j| = 2^{\tau}$}
		\STATE $\Lambda_j(x) = s_{\tau}(x) - s_{\tau}(\omega_{j\cdot 2^{\tau}})$.
		\ELSE
		\STATE Compute $R_{\omega_{j\cdot 2^{\tau} + l}}$ and obtain $T_l$, $l = 0, 1, \dots, 2^{\tau}$.
		\STATE Compute $\mathbf{\hat{T}} = \text{FWHT}(\mathbf{T})$.
		\STATE Compute the component-wise multiplication between $\mathbf{\hat{T}}$ and $\mathbf{\hat{L}}$ and obtain $\mathbf{\hat{W}}$.
		\STATE Perform a $2^{\tau}$-point inverse FWHT on $\mathbf{\hat{W}}$ to obtain $\log(\Lambda_j(\omega_{j\cdot 2^{\tau} + l}))$, $l = 0, 1, \dots, 2^{\tau} - 1$.
		\STATE Compute $\Lambda_j(\omega_{j\cdot 2^{\tau} + l})$, $l = 0, 1, \dots, 2^{\tau} - 1$.
		\STATE Set $\Lambda_j(\omega_{j\cdot 2^{\tau} + l}) = 0$ for all $\omega_{j\cdot 2^{\tau} + l}\in E_j$.
		\STATE Perform a $2^{\tau}$-point IFFT to obtain $\Lambda_j(x)$.
		\ENDIF
	\end{algorithmic}
\end{algorithm}

We now present the method of computing $\Lambda(x)$.

Given $E$, we first find the desired $\tau$ such that $2^{\tau}\geq|E|$.
Clearly, we have
\begin{equation*}
	E = \bigcup_{j = 0}^{\eta - 1}E_j, \Lambda(x) = \prod_{j = 0}^{\eta - 1}\Lambda_j(x).
\end{equation*}

For each subset $E_j$, the corresponding $\Lambda_j(x)$ can be computed by Algorithm \ref{alg:coset}. We evaluate each $\Lambda_j(x)$ at the set $P_0$ by a $2^{\tau}$-point FFT and at the element $\omega_{2^{\tau}}$ by substituting it. We then obtain the evaluations
\begin{equation*}
	\Lambda_j(\omega_0), \Lambda_j(\omega_1), \dots, \Lambda_j(\omega_{2^{\tau} - 1}), \Lambda_j(\omega_{2^{\tau}}),
\end{equation*}
for all possible $j$.
Next, we compute the evaluation of $\Lambda(x)$ by
\begin{equation}\label{eq:eval_prod}
	\Lambda(\omega_l) = \prod_{j = 0}^{\eta - 1}\Lambda_j(\omega_l), l = 0, 1, \dots, 2^{\tau}.
\end{equation}
Finally, given the evaluations $\Lambda(\omega_0), \Lambda(\omega_1), \dots, \Lambda(\omega_{2^{\tau}})$, we can obtain $\Lambda(x)$ by an extended $2^{\tau}$-point IFFT algorithm, see \cite[Algorithm 5]{tang2022a}.

Algorithm \ref{alg:total} describe the method of computing $\Lambda(x)$. 
%该算法的计算复杂度为
%\begin{equation*}
%	O(2^{\tau}\log(2^{\tau})) \times \eta = O(2^m\log(2^{\tau})).
%\end{equation*}

\begin{algorithm}[h]
	\caption{Computing $\Lambda(x)$}
	\label{alg:total}
	\begin{algorithmic}[1]
		\REQUIRE $E$.
		\ENSURE $\Lambda(x)$.
		\STATE Compute $\tau$ and $E_j$, $j = 0, 1, \dots, \eta - 1$.
		\STATE Compute $\Lambda_j(x)$ by Algorithm \ref{alg:coset} for $j = 0, 1, \dots, \eta - 1$.
		\STATE Evaluate $\Lambda_j(x)$ at $P_0$ and $\omega_{2^{\tau}}$ for $j = 0, 1, \dots, \eta - 1$.
		\STATE Compute the evaluations of $\Lambda(x)$ by \eqref{eq:eval_prod}.
		\STATE Perform an extended $2^{\tau}$-point IFFT to obtain $\Lambda(x)$.
	\end{algorithmic}
\end{algorithm}

We now analyze the computational complexity of Algorithm \ref{alg:total}. The complexity of Step 1 is $O(n)$. Step 2 needs $\eta \times O(2^{\tau}\log(2^{\tau})) = O(2^m\log(2^{\tau}))$ operations. Step 3 also needs $\eta \times O(2^{\tau}\log(2^{\tau})) = O(2^m\log(2^{\tau}))$ operations. Finally, Step 4 requires $(2^{\tau} + 1)\times \eta = O(2^m)$ operations and Step 5 requires $O(2^{\tau}\log(2^{\tau}))$ operations. Hence, one can conclude that the complexity of Algorithm \ref{alg:total} is $O(2^m\log(2^{\tau}))$.

\section{Conclusion}\label{sec:conclusion}
This paper proposes an improved algorithm for computing vanishing polynomials, which achieves the best known complexity to date. With this algorithm, the complexity of erasure decoding for high-rate RS codes is reduced to $O(n\log(n - k))$, where $n$ is the code length and $k$ is the code dimension. This improves upon the complexity bound in \cite{Chen2026Two}.

%% The very first letter is a 2 line initial drop letter followed
%% by the rest of the first word in caps.
%% 
%% form to use if the first word consists of a single letter:
%% \IEEEPARstart{A}{demo} file is ....
%% 
%% form to use if you need the single drop letter followed by
%% normal text (unknown if ever used by the IEEE):
%% \IEEEPARstart{A}{}demo file is ....
%% 
%% Some journals put the first two words in caps:
%% \IEEEPARstart{T}{his demo} file is ....
%% 
%% Here we have the typical use of a "T" for an initial drop letter
%% and "HIS" in caps to complete the first word.

\ifCLASSOPTIONcaptionsoff
  \newpage
\fi
\bibliographystyle{IEEEtran}
\bibliography{refs_final}

\end{document}